\documentclass[conference]{IEEEtran}

\usepackage{cite}
\usepackage{amsmath,amssymb,amsfonts}
\usepackage{algorithmic}
\usepackage{algorithm}
\usepackage{graphicx}
\usepackage{textcomp}
\usepackage{xcolor}
\usepackage{booktabs}
\usepackage{multirow}
\usepackage{hyperref}
\usepackage{balance}
\usepackage{enumitem}
\usepackage{subcaption}
\usepackage{array}

\hypersetup{
    colorlinks=true,
    linkcolor=black,
    citecolor=black,
    urlcolor=blue
}

\def\BibTeX{{\rm B\kern-.05em{\sc i\kern-.025em b}\kern-.08em
    T\kern-.1667em\lower.7ex\hbox{E}\kern-.125emX}}

\begin{document}

\title{CCCE: A Continuous Code Calibration Engine for Autonomous Enterprise Codebase Maintenance via Knowledge Graph Traversal and Adaptive Decision Gating}

\author{
\IEEEauthorblockN{Santhosh Kusuma Kumar Parimi}
\IEEEauthorblockA{Independent Researcher \\ Austin, TX, USA \\
santuparimi86@gmail.com}
}

\maketitle

\begin{abstract}
Enterprise software organizations face an escalating challenge in maintaining the integrity, security, and freshness of codebases that span hundreds of repositories, multiple programming languages, and thousands of interdependent packages. Existing approaches to codebase maintenance---including static analysis, software composition analysis (SCA), and dependency management tools---operate in isolation, address only narrow subsets of maintenance concerns, and require substantial manual intervention to propagate changes across interconnected systems. We present the \textbf{Continuous Code Calibration Engine (CCCE)}, an event-driven, AI-agentic system that autonomously maintains enterprise codebases throughout the Software Development Life Cycle (SDLC). The CCCE introduces three key technical innovations: (1)~a dynamic knowledge graph with bidirectional traversal algorithms that simultaneously compute forward impact propagation and backward test adequacy analysis; (2)~an adaptive multi-stage gating framework that classifies calibration actions into four risk tiers using learned risk-confidence scoring rather than static rules; and (3)~a multi-model continuous learning architecture operating at multiple temporal scales to refine calibration strategies, risk models, and organizational policies from operational feedback. We formalize the system's graph model, traversal algorithms, and decision logic, and demonstrate through three representative enterprise scenarios that the CCCE reduces mean time to remediation by enabling coordinated, cross-repository calibrations with human-in-the-loop (HITL) oversight where appropriate. The system generates atomic, semantically verified patches with progressive validation and intelligent rollback capabilities, providing end-to-end traceability from triggering events through calibration execution and outcome learning.
\end{abstract}


\section{Introduction}

Modern enterprise software ecosystems are characterized by extraordinary complexity: hundreds to thousands of repositories, polyglot technology stacks spanning multiple languages and frameworks, and intricate webs of dependencies on external packages, internal libraries, and shared services~\cite{decan2019empirical}. This complexity creates a fundamental maintenance challenge---when a software package releases a new version, an API is deprecated, or a Common Vulnerabilities and Exposures (CVE) advisory is disclosed, the impact can propagate across dozens or hundreds of interconnected projects in ways that are difficult to predict, trace, or remediate~\cite{pashchenko2020vuln4real}.

The current state of practice relies on a fragmented toolchain: static analysis tools such as SonarQube identify code quality issues but do not remediate them~\cite{vassallo2020developers}; software composition analysis (SCA) tools such as Snyk and OWASP Dependency-Check detect vulnerable dependencies but typically operate on individual projects without cross-repository impact analysis~\cite{plate2015impact}; dependency management tools automate package resolution but lack security and quality intelligence~\cite{kikas2017structure}; and CI/CD pipelines orchestrate builds and deployments but provide no mechanism for intelligent, automated code calibration~\cite{shahin2017continuous}. This fragmentation leads to several critical problems:

\begin{enumerate}[leftmargin=*]
    \item \textbf{Fragmented visibility}: Organizations lack a holistic view of interdependencies across repositories, languages, and tooling, making comprehensive impact analysis infeasible~\cite{zimmermann2019small}.
    \item \textbf{Reactive maintenance}: Current workflows respond to issues after they manifest rather than proactively detecting and addressing impending risks, leading to dependency drift and accumulated technical debt~\cite{besker2018managing}.
    \item \textbf{Inconsistent propagation}: CVE disclosures and dependency updates are not systematically propagated to all affected projects, leaving organizations with partial remediation and residual risk~\cite{pashchenko2020vuln4real}.
    \item \textbf{Manual remediation burden}: Even when issues are identified, the process of planning, implementing, testing, and deploying fixes across multiple repositories remains largely manual, consuming significant engineering effort~\cite{li2020understanding}.
    \item \textbf{Weak auditability}: The lack of end-to-end traceability from triggering events through remediation actions to outcomes impedes compliance and governance requirements~\cite{ruohonen2018empirical}.
\end{enumerate}

To address these challenges, we present the \textbf{Continuous Code Calibration Engine (CCCE)}, an autonomous, event-driven system that provides end-to-end codebase maintenance through the integration of knowledge graph reasoning, adaptive decision gating, and continuous learning. The key contributions of this paper are as follows:

\begin{itemize}[leftmargin=*]
    \item A \textbf{dynamic knowledge graph model} with typed nodes and relationships that captures the enterprise software ecosystem, including projects, packages, CVEs, APIs, and tests, along with computed attributes such as impact radius, change propagation risk, and remediation cost.
    \item \textbf{Bidirectional traversal algorithms} that simultaneously compute forward impact propagation through dependency chains and backward test adequacy analysis, yielding prioritized remediation plans rather than mere vulnerability reports.
    \item An \textbf{adaptive multi-stage gating framework} that classifies calibrations into four action types (Automated Safe, Automated with Validation, Human-Assisted, and Advisory Only) using dynamic risk-confidence scoring that learns from operational outcomes.
    \item A \textbf{multi-model continuous learning architecture} with four specialized models operating at different temporal scales (continuous, weekly, bi-weekly, monthly) that refine calibration strategies, risk scoring, test adequacy assessment, and code transformation patterns.
    \item \textbf{Atomic patch generation with semantic preservation verification} that ensures API contract compatibility and enables fine-grained rollback at individual change unit boundaries.
\end{itemize}

The remainder of this paper is organized as follows. Section~\ref{sec:related} surveys related work. Section~\ref{sec:architecture} presents the system architecture. Section~\ref{sec:knowledge_graph} details the knowledge graph model and traversal algorithms. Section~\ref{sec:calibration} describes the calibration decision framework. Section~\ref{sec:learning} presents the learning and adaptation mechanisms. Section~\ref{sec:evaluation} provides case study evaluations. Section~\ref{sec:discussion} discusses implications and limitations, and Section~\ref{sec:conclusion} concludes.

\section{Related Work}
\label{sec:related}

The CCCE draws on and advances several established research areas. We organize related work into five categories corresponding to the system's key capabilities.

\subsection{Software Composition Analysis and Vulnerability Management}

Software Composition Analysis (SCA) tools identify known vulnerabilities in third-party dependencies by matching library versions against CVE databases~\cite{plate2015impact}. Tools such as OWASP Dependency-Check, Snyk, and Mend (formerly WhiteSource) have become standard in modern development workflows~\cite{pashchenko2020vuln4real}. However, these tools primarily operate at the individual project level and focus on \textit{detection} rather than \textit{remediation}. Plate et al.~\cite{plate2015impact} demonstrated that vulnerability detection alone is insufficient---organizations need to understand whether vulnerable code paths are actually reachable. The CCCE extends SCA capabilities by not only detecting vulnerabilities but also computing cross-repository impact through knowledge graph traversal and automatically generating remediation patches.

\subsection{Dependency Management and Analysis}

Research on software dependency networks has revealed the complex, fragile structure of modern package ecosystems~\cite{kikas2017structure, decan2019empirical}. Kikas et al.~\cite{kikas2017structure} analyzed dependency networks across npm, RubyGems, and Rust, finding that transitive dependencies significantly amplify risk. Zimmermann et al.~\cite{zimmermann2019small} demonstrated that compromising a single popular npm package can affect thousands of downstream projects. Tools like Dependabot and Renovate automate dependency version updates but operate on individual repositories without cross-project coordination or semantic verification of code compatibility~\cite{alfadel2021empirical}. The CCCE addresses these limitations through its knowledge graph model that captures transitive dependencies and enables coordinated, enterprise-wide calibration.

\subsection{Automated Program Repair and Code Transformation}

Automated program repair (APR) techniques generate patches for software defects through search-based~\cite{le2012genprog}, semantics-based~\cite{nguyen2013semfix}, and learning-based approaches~\cite{chen2019sequencer}. Recent work has explored large language models for code repair~\cite{xia2023automated}, demonstrating promising results on benchmark datasets. However, APR research has primarily focused on bug-fixing in isolated codebases rather than enterprise-wide maintenance tasks such as dependency updates and API migration. The CCCE's patch generation mechanism differs fundamentally: it produces atomic, semantically verified patches specifically designed for calibration tasks (dependency updates, API migrations, CVE remediation) with embedded traceability metadata.

\subsection{Knowledge Graphs in Software Engineering}

Knowledge graphs have been applied to software engineering for tasks including code search~\cite{gu2018deep}, API recommendation~\cite{chen2021holistic}, and vulnerability analysis~\cite{zheng2021d2a}. Zheng et al.~\cite{zheng2021d2a} constructed vulnerability knowledge graphs to assist in vulnerability assessment, while Chen et al.~\cite{chen2021holistic} used knowledge graphs for holistic API recommendation. The CCCE's knowledge graph is distinctive in its scope (encompassing projects, packages, CVEs, APIs, and tests in a unified model) and its use for \textit{operational} purposes---driving automated calibration decisions rather than serving as a passive information resource.

\subsection{CI/CD and DevOps Automation}

Modern CI/CD practices have significantly improved software delivery speed and reliability~\cite{shahin2017continuous}. However, CI/CD pipelines remain primarily \textit{orchestration} engines that execute predefined workflows rather than \textit{intelligent} systems that reason about what actions to take~\cite{leite2019survey}. Recent work on AIOps has explored applying machine learning to operations~\cite{dang2019aiops}, but this research focuses on monitoring and incident response rather than proactive codebase maintenance. The CCCE complements existing CI/CD infrastructure by providing the intelligence layer for automated, context-aware code calibration.

\begin{table*}[!t]
\centering
\caption{Comparison of CCCE with Existing Approaches Across Key Capability Dimensions}
\label{tab:comparison}
\renewcommand{\arraystretch}{1.2}
\begin{tabular}{p{3.8cm}ccccccc}
\toprule
\textbf{Capability} & \textbf{SCA} & \textbf{Static} & \textbf{Dep.\ Mgmt.} & \textbf{APR} & \textbf{CI/CD} & \textbf{Vuln.\ Mgmt.} & \textbf{CCCE} \\
 & \textbf{Tools} & \textbf{Analysis} & \textbf{Tools} & \textbf{Tools} & \textbf{Pipelines} & \textbf{Platforms} & \textbf{(Ours)} \\
\midrule
Cross-repo impact analysis & \textcolor{red}{$\times$} & \textcolor{red}{$\times$} & \textcolor{red}{$\times$} & \textcolor{red}{$\times$} & \textcolor{red}{$\times$} & Partial & \textcolor{green!50!black}{\checkmark} \\
Knowledge graph reasoning & \textcolor{red}{$\times$} & \textcolor{red}{$\times$} & \textcolor{red}{$\times$} & \textcolor{red}{$\times$} & \textcolor{red}{$\times$} & \textcolor{red}{$\times$} & \textcolor{green!50!black}{\checkmark} \\
Automated code calibration & \textcolor{red}{$\times$} & \textcolor{red}{$\times$} & Partial & Partial & \textcolor{red}{$\times$} & \textcolor{red}{$\times$} & \textcolor{green!50!black}{\checkmark} \\
Adaptive risk-based gating & \textcolor{red}{$\times$} & \textcolor{red}{$\times$} & \textcolor{red}{$\times$} & \textcolor{red}{$\times$} & Static & \textcolor{red}{$\times$} & \textcolor{green!50!black}{\checkmark} \\
Semantic patch verification & \textcolor{red}{$\times$} & Partial & \textcolor{red}{$\times$} & \textcolor{red}{$\times$} & \textcolor{red}{$\times$} & \textcolor{red}{$\times$} & \textcolor{green!50!black}{\checkmark} \\
Continuous learning & \textcolor{red}{$\times$} & \textcolor{red}{$\times$} & \textcolor{red}{$\times$} & \textcolor{red}{$\times$} & \textcolor{red}{$\times$} & \textcolor{red}{$\times$} & \textcolor{green!50!black}{\checkmark} \\
HITL-selective oversight & \textcolor{red}{$\times$} & \textcolor{red}{$\times$} & \textcolor{red}{$\times$} & \textcolor{red}{$\times$} & Manual & \textcolor{red}{$\times$} & \textcolor{green!50!black}{\checkmark} \\
End-to-end traceability & \textcolor{red}{$\times$} & \textcolor{red}{$\times$} & \textcolor{red}{$\times$} & \textcolor{red}{$\times$} & Partial & Partial & \textcolor{green!50!black}{\checkmark} \\
\bottomrule
\end{tabular}
\end{table*}

Table~\ref{tab:comparison} summarizes the capability comparison between existing approaches and the CCCE. The key distinction is that the CCCE integrates all of these capabilities into a unified, event-driven architecture with continuous learning, whereas existing solutions address individual dimensions in isolation.

\section{System Architecture}
\label{sec:architecture}

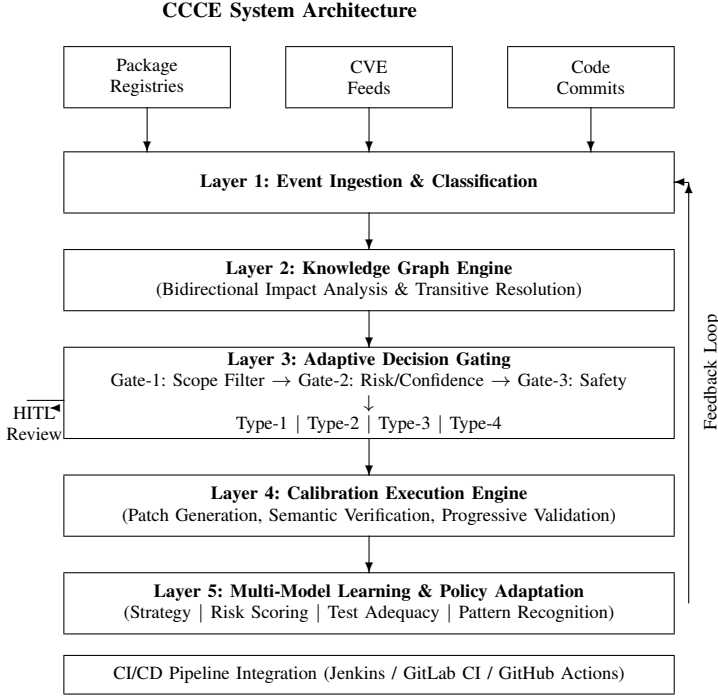
\begin{figure}[!t]
\centering
\setlength{\unitlength}{1cm}
\begin{picture}(8.5,11.5)
\put(1.5,11){\footnotesize\textbf{CCCE System Architecture}}

\put(0.2,9.8){\framebox(2.2,0.8){\parbox{2cm}{\centering\scriptsize Package\\Registries}}}
\put(3.15,9.8){\framebox(2.2,0.8){\parbox{2cm}{\centering\scriptsize CVE\\Feeds}}}
\put(6.1,9.8){\framebox(2.2,0.8){\parbox{2cm}{\centering\scriptsize Code\\Commits}}}

\put(1.3,9.8){\vector(0,-1){0.6}}
\put(4.25,9.8){\vector(0,-1){0.6}}
\put(7.2,9.8){\vector(0,-1){0.6}}

\put(0.2,8.4){\framebox(8.1,0.8){\parbox{7.8cm}{\centering\scriptsize\textbf{Layer 1: Event Ingestion \& Classification}}}}

\put(4.25,8.4){\vector(0,-1){0.5}}

\put(0.2,7.1){\framebox(8.1,0.8){\parbox{7.8cm}{\centering\scriptsize\textbf{Layer 2: Knowledge Graph Engine}\\(Bidirectional Impact Analysis \& Transitive Resolution)}}}

\put(4.25,7.1){\vector(0,-1){0.5}}

\put(0.2,5.4){\framebox(8.1,1.2){\parbox{7.8cm}{\centering\scriptsize\textbf{Layer 3: Adaptive Decision Gating}\\Gate-1: Scope Filter $\rightarrow$ Gate-2: Risk/Confidence $\rightarrow$ Gate-3: Safety\\$\downarrow$\\Type-1 $|$ Type-2 $|$ Type-3 $|$ Type-4}}}

\put(4.25,5.4){\vector(0,-1){0.5}}

\put(0.2,4.1){\framebox(8.1,0.8){\parbox{7.8cm}{\centering\scriptsize\textbf{Layer 4: Calibration Execution Engine}\\(Patch Generation, Semantic Verification, Progressive Validation)}}}

\put(4.25,4.1){\vector(0,-1){0.5}}

\put(0.2,2.8){\framebox(8.1,0.8){\parbox{7.8cm}{\centering\scriptsize\textbf{Layer 5: Multi-Model Learning \& Policy Adaptation}\\(Strategy $|$ Risk Scoring $|$ Test Adequacy $|$ Pattern Recognition)}}}

\put(8.5,3.2){\vector(0,1){5.6}}
\put(8.5,8.8){\vector(-1,0){0.2}}
\put(8.7,5.5){\rotatebox{90}{\scriptsize Feedback Loop}}

\put(0.0,5.8){\vector(-1,0){0.0}}
\put(-0.6,5.5){\parbox{0.8cm}{\centering\scriptsize HITL\\Review}}
\put(0.2,5.9){\line(-1,0){0.5}}

\put(0.2,2.0){\framebox(8.1,0.5){\parbox{7.8cm}{\centering\scriptsize CI/CD Pipeline Integration (Jenkins / GitLab CI / GitHub Actions)}}}

\end{picture}
\caption{High-level architecture of the CCCE showing the five-layer processing pipeline with continuous feedback loop, HITL integration, and CI/CD pipeline connectivity.}
\label{fig:architecture}
\end{figure}

The CCCE is architected as an event-driven, multi-layer system comprising five principal components that operate in a continuous feedback loop (Fig.~\ref{fig:architecture}).

\textbf{Layer 1: Event Ingestion and Classification.} The system subscribes to heterogeneous event sources including package registry feeds (npm, Maven Central, PyPI), CVE advisory databases (NVD, GitHub Security Advisories), code repository webhooks, and API specification change notifications. Events are normalized into a canonical format and classified by type (package update, CVE disclosure, API deprecation, configuration change) and preliminary severity.

\textbf{Layer 2: Knowledge Graph Engine.} Upon receiving a classified event, the knowledge graph engine identifies the corresponding root nodes and executes the bidirectional impact analysis algorithm (Section~\ref{sec:knowledge_graph}). The output is a prioritized set of affected projects with associated impact scores, test adequacy assessments, and recommended calibration priorities.

\textbf{Layer 3: Adaptive Decision Gating.} For each affected project, the multi-stage gating framework (Section~\ref{sec:calibration}) classifies the required calibration into one of four action types based on dynamic risk-confidence scoring. This classification determines the degree of automation and human oversight applied.

\textbf{Layer 4: Calibration Execution Engine.} Based on the action classification, the execution engine generates atomic, semantically verified patches; executes progressive validation (syntax, unit tests, integration tests, contract tests, performance benchmarks); and applies intelligent rollback logic when validation failures occur.

\textbf{Layer 5: Multi-Model Learning and Policy Adaptation.} Calibration outcomes, human feedback, and business metrics are continuously collected and fed into four specialized learning models (Section~\ref{sec:learning}) that refine the system's calibration strategies, risk scoring, test adequacy assessment, and code transformation patterns.

\section{Knowledge Graph Model and Traversal Algorithms}
\label{sec:knowledge_graph}

\subsection{Graph Schema}

The CCCE constructs and maintains a dynamic knowledge graph $G = (V, E, \phi, \psi)$ where $V$ is the set of typed nodes, $E$ is the set of typed edges, $\phi: V \rightarrow \mathcal{A}_V$ maps nodes to their attribute sets, and $\psi: E \rightarrow \mathcal{A}_E$ maps edges to their attribute sets.

\subsubsection{Node Types}

The graph comprises five node types, each with domain-specific attributes:

\begin{itemize}[leftmargin=*]
    \item \textbf{PROJECT} ($v_p$): Represents a code repository or service, with attributes: repository identifier $\text{id}_r$, programming language $\lambda$, business criticality $c \in \{\text{critical}, \text{high}, \text{medium}, \text{low}\}$, and test coverage percentage $\tau \in [0, 1]$.
    \item \textbf{PACKAGE} ($v_k$): Represents an external library, with attributes: package identifier $\text{id}_k$, version $\nu$, ecosystem $\epsilon \in \{\text{npm}, \text{maven}, \text{pip}, \ldots\}$, and release timestamp $t_r$.
    \item \textbf{CVE} ($v_c$): Represents a disclosed vulnerability, with attributes: CVE identifier $\text{id}_c$, CVSS score $s_{\text{cvss}} \in [0, 10]$, severity class $\sigma$, and disclosure date $t_d$.
    \item \textbf{API} ($v_a$): Represents an exposed or consumed interface, with attributes: API signature $\text{sig}_a$, deprecation flag $\delta \in \{0, 1\}$, and usage frequency $f_u$.
    \item \textbf{TEST} ($v_t$): Represents a test suite or test case, with attributes: test identifier $\text{id}_t$, scope coverage $\omega$, and historical pass rate $\rho \in [0, 1]$.
\end{itemize}

\subsubsection{Relationship Types}

Five relationship types capture the structural and semantic connections:

\begin{itemize}[leftmargin=*]
    \item $\textsc{depends\_on}(v_p, v_k)$: Links projects to packages, annotated with dependency depth $d \in \{\text{direct}, \text{transitive}\}$ and dependency strength $w_d \in [0, 1]$.
    \item $\textsc{exposes}(v_p, v_a)$ / $\textsc{consumes}(v_p, v_a)$: Bidirectional links between projects and APIs, annotated with coupling strength $w_c \in [0, 1]$.
    \item $\textsc{affects}(v_c, v_k)$: Links CVEs to vulnerable packages, annotated with exploitability rating $e \in [0, 1]$.
    \item $\textsc{tests}(v_t, v_p)$: Links tests to projects, annotated with coverage percentage $\tau_t \in [0, 1]$.
    \item $\textsc{trans\_depends}(v_p, v_k)$: Computed transitive dependency through multi-hop paths.
\end{itemize}

\subsubsection{Computed Attributes}

Each node and relationship maintains dynamically computed metrics:
\begin{itemize}[leftmargin=*]
    \item \textbf{Impact radius} $\mathcal{I}(v)$: The number of downstream dependents weighted by their criticality.
    \item \textbf{Change propagation risk} $\mathcal{R}(e)$: Estimated likelihood that upstream changes cause downstream failures.
    \item \textbf{Remediation cost} $\mathcal{C}(v)$: Estimated effort to calibrate affected code.
\end{itemize}

\subsection{Bidirectional Impact Analysis}

Algorithm~\ref{alg:bidi} presents the bidirectional impact analysis procedure. Given a triggering event, it simultaneously computes forward impact through dependency chains and backward test adequacy, yielding a prioritized remediation plan.

\begin{algorithm}[!t]
\caption{Bidirectional Impact Analysis}
\label{alg:bidi}
\begin{algorithmic}[1]
\REQUIRE Event $\mathcal{E}$, Knowledge Graph $G$, Threshold $\theta$
\ENSURE Prioritized affected project set $\mathcal{P}^*$
\STATE $V_{\text{root}} \leftarrow \text{IdentifyRootNodes}(G, \mathcal{E})$
\STATE $\mathcal{P}_{\text{affected}} \leftarrow \emptyset$
\STATE $Q \leftarrow V_{\text{root}}$ \COMMENT{BFS queue}
\FORALL{$v_r \in V_{\text{root}}$}
    \STATE $\text{depth}(v_r) \leftarrow 0$
\ENDFOR
\WHILE{$Q \neq \emptyset$}
    \STATE $v \leftarrow Q.\text{dequeue}()$
    \FORALL{$v_p \in \text{Neighbors}(v, \textsc{depends\_on})$}
        \STATE $s \leftarrow \text{Severity}(\mathcal{E}) \times \alpha^{\text{depth}(v)+1} \times w(c(v_p))$
        \IF{$s > \theta$}
            \STATE $\mathcal{P}_{\text{affected}} \leftarrow \mathcal{P}_{\text{affected}} \cup \{(v_p, s)\}$
            \STATE $\text{depth}(v_p) \leftarrow \text{depth}(v) + 1$
            \STATE $Q.\text{enqueue}(v_p)$
        \ENDIF
    \ENDFOR
\ENDWHILE
\FORALL{$(v_p, s) \in \mathcal{P}_{\text{affected}}$}
    \STATE $\tau_p \leftarrow \text{TestAdequacy}(G, v_p)$ \COMMENT{Backward traversal}
    \STATE $\text{priority}(v_p) \leftarrow w_1 \cdot s + w_2 \cdot \mathcal{C}(v_p) + w_3 \cdot (1 - \tau_p)$
\ENDFOR
\STATE $\mathcal{P}^* \leftarrow \text{SortByPriority}(\mathcal{P}_{\text{affected}})$
\RETURN $\mathcal{P}^*$
\end{algorithmic}
\end{algorithm}

The impact score for a project $v_p$ with respect to event $\mathcal{E}$ is computed as:

\begin{equation}
\label{eq:impact}
S_{\text{impact}}(v_p, \mathcal{E}) = \sum_{\pi \in \Pi(v_{\mathcal{E}}, v_p)} S_{\text{sev}}(\mathcal{E}) \cdot \alpha^{|\pi|} \cdot w(c(v_p))
\end{equation}

\noindent where $\Pi(v_{\mathcal{E}}, v_p)$ is the set of all paths from the event source node to $v_p$, $S_{\text{sev}}(\mathcal{E})$ is the normalized event severity (e.g., CVSS/10 for CVEs), $\alpha \in (0, 1)$ is the distance decay factor, $|\pi|$ is the path length, and $w(c(v_p))$ is the criticality weight of the project.

\subsection{Transitive Dependency Resolution}

Algorithm~\ref{alg:transitive} computes the complete dependency closure for any project, annotating each dependency with relevant security metadata.

\begin{algorithm}[!t]
\caption{Transitive Dependency Resolution with CVE Annotation}
\label{alg:transitive}
\begin{algorithmic}[1]
\REQUIRE Project node $v_p$, Graph $G$, Depth limit $D_{\max}$
\ENSURE Annotated dependency closure $\mathcal{D}$
\STATE $\mathcal{D} \leftarrow \emptyset$, $Q \leftarrow \{(v_p, 0)\}$
\WHILE{$Q \neq \emptyset$ \AND $d < D_{\max}$}
    \STATE $(v, d) \leftarrow Q.\text{dequeue}()$
    \FORALL{$v_k \in \text{Neighbors}(v, \textsc{depends\_on})$}
        \IF{$v_k \notin \mathcal{D}$}
            \STATE $\mathcal{V}_c \leftarrow \{v_c : (v_c, v_k) \in E_{\textsc{affects}}\}$
            \STATE $\mathcal{D} \leftarrow \mathcal{D} \cup \{(v_k, d+1, \mathcal{V}_c)\}$
            \STATE $Q.\text{enqueue}(v_k, d+1)$
        \ENDIF
    \ENDFOR
\ENDWHILE
\RETURN $\mathcal{D}$
\end{algorithmic}
\end{algorithm}

\subsection{Graph Maintenance}

The knowledge graph is continuously maintained through four mechanisms: (1)~\textit{event stream integration} for real-time ingestion of package releases, CVE disclosures, and code commits; (2)~\textit{dependency scanning} via periodic analysis of project manifests (e.g., \texttt{package.json}, \texttt{pom.xml}, \texttt{requirements.txt}); (3)~\textit{API discovery} through static and dynamic analysis to identify usage patterns; and (4)~\textit{decay and pruning} to remove outdated nodes and relationships based on configurable retention policies.

\section{Calibration Decision Scoring and Gating Framework}
\label{sec:calibration}

\subsection{Calibration Action Classification}

Each identified calibration need is classified into one of four tiers based on complexity, risk, and the degree of human oversight required:

\begin{itemize}[leftmargin=*]
    \item \textbf{Type-1 (Automated Safe)}: Low-risk changes with high-confidence automation. Examples include version bumps in manifest files, non-breaking dependency updates, and documentation updates. \textit{Characteristics}: No code logic changes, backward compatible, fully reversible.
    \item \textbf{Type-2 (Automated with Validation)}: Moderate changes requiring automated testing. Examples include API signature updates with well-defined refactoring rules and deprecated method replacements with direct equivalents. \textit{Characteristics}: Code logic affected but transformation rules are well-established.
    \item \textbf{Type-3 (Human-Assisted)}: Complex changes requiring human review before commit. Examples include breaking API changes, security-critical authentication logic updates, and multi-service coordination changes. \textit{Characteristics}: Semantic complexity with potential for cascading failures.
    \item \textbf{Type-4 (Advisory Only)}: Changes requiring human design decisions. Examples include major version upgrades with architectural implications and API deprecations without clear migration paths. \textit{Characteristics}: Strategic nature with multiple valid solution approaches.
\end{itemize}

\subsection{Multi-Dimensional Scoring Model}

For each affected project, the system computes a \textit{risk score} $R$ and a \textit{confidence score} $C$ from orthogonal component dimensions.

\subsubsection{Risk Score}

The risk score aggregates six components:

\begin{equation}
\label{eq:risk}
R = \sum_{i=1}^{6} w_i^R \cdot r_i, \quad \text{where } \sum_{i=1}^{6} w_i^R = 1
\end{equation}

\noindent where the components $r_i$ (each normalized to $[0, 1]$) are: (1)~change scope (lines, files, modules affected), (2)~blast radius (number of dependent projects), (3)~semantic complexity (degree of business logic involvement), (4)~test coverage gap ($1 - \tau$), (5)~historical instability (past failure frequency), and (6)~criticality inheritance (business criticality of the project and downstream dependents).

\subsubsection{Confidence Score}

The confidence score aggregates four components:

\begin{equation}
\label{eq:confidence}
C = \sum_{j=1}^{4} w_j^C \cdot c_j, \quad \text{where } \sum_{j=1}^{4} w_j^C = 1
\end{equation}

\noindent where the components $c_j$ (each normalized to $[0, 1]$) are: (1)~pattern recognition match (similarity to previously successful calibrations), (2)~static analysis certainty, (3)~test suite strength (comprehensiveness of available tests), and (4)~transformation rule maturity (whether the required transformation follows well-established patterns).

\subsection{Three-Gate Decision Logic}

The system applies a cascading three-gate decision tree (Fig.~\ref{fig:gating}) to classify each calibration:

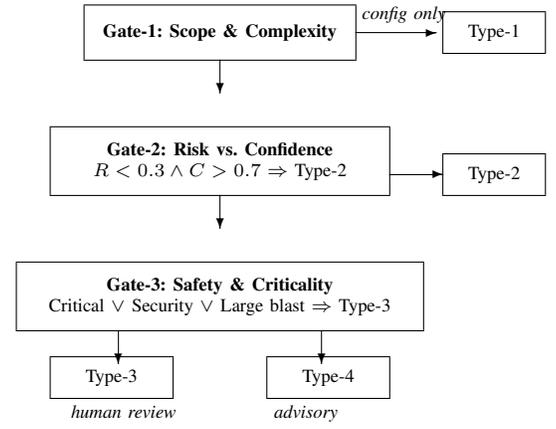
\begin{figure}[!t]
\centering
\setlength{\unitlength}{0.9cm}
\begin{picture}(9,9.5)
\put(2.5,8.5){\framebox(4,0.8){\scriptsize\textbf{Gate-1: Scope \& Complexity}}}
\put(4.5,8.5){\vector(0,-1){0.5}}
\put(6.5,8.9){\vector(1,0){1.2}}
\put(7.8,8.6){\framebox(1.5,0.6){\scriptsize Type-1}}
\put(6.6,9.1){\scriptsize\textit{config only}}

\put(2,6.5){\framebox(5,1){\parbox{4.6cm}{\centering\scriptsize\textbf{Gate-2: Risk vs.\ Confidence}\\\scriptsize $R < 0.3 \land C > 0.7 \Rightarrow$ Type-2}}}
\put(4.5,6.5){\vector(0,-1){0.5}}
\put(7,6.8){\vector(1,0){0.8}}
\put(7.8,6.5){\framebox(1.5,0.6){\scriptsize Type-2}}

\put(1.5,4.5){\framebox(6,1){\parbox{5.6cm}{\centering\scriptsize\textbf{Gate-3: Safety \& Criticality}\\\scriptsize Critical $\lor$ Security $\lor$ Large blast $\Rightarrow$ Type-3}}}
\put(3,4.5){\vector(0,-1){0.5}}
\put(6,4.5){\vector(0,-1){0.5}}
\put(2,3.5){\framebox(1.8,0.6){\scriptsize Type-3}}
\put(5.2,3.5){\framebox(1.8,0.6){\scriptsize Type-4}}
\put(2.3,3.2){\scriptsize\textit{human review}}
\put(5.3,3.2){\scriptsize\textit{advisory}}

\end{picture}
\caption{Three-gate decision logic for calibration action classification. Events cascade through gates until classified into one of four action types.}
\label{fig:gating}
\end{figure}

\textbf{Gate-1 (Scope and Complexity Filter):} If the change is limited to configuration or manifest files, affects no code logic, and is backward compatible, it is classified as Type-1.

\textbf{Gate-2 (Risk vs.\ Confidence Assessment):} The risk and confidence scores are evaluated against configurable thresholds:
\begin{itemize}[leftmargin=*]
    \item If $R < \theta_R^{\text{low}}$ and $C > \theta_C^{\text{high}}$: classify as Type-2
    \item If $R < \theta_R^{\text{mod}}$ and $C > \theta_C^{\text{high}}$ and $\tau > \theta_\tau$: classify as Type-2
    \item Otherwise: proceed to Gate-3
\end{itemize}

\noindent where default thresholds are $\theta_R^{\text{low}} = 0.3$, $\theta_R^{\text{mod}} = 0.6$, $\theta_C^{\text{high}} = 0.7$, and $\theta_\tau = 0.75$.

\textbf{Gate-3 (Safety and Criticality Assessment):} Evaluates project criticality, security-criticality flags, and blast radius:
\begin{itemize}[leftmargin=*]
    \item If criticality is \textit{critical}, security flag is set, or blast radius exceeds 10 dependent services: classify as Type-3
    \item If architectural decisions are required or no clear transformation path exists: classify as Type-4
    \item Otherwise: classify as Type-3 as conservative default
\end{itemize}

\subsection{Calibration Strategy Execution}

Each action type triggers a specific execution strategy:

\begin{table}[!t]
\centering
\caption{Calibration Strategy by Action Type}
\label{tab:strategies}
\renewcommand{\arraystretch}{1.15}
\begin{tabular}{p{1.1cm}p{2.2cm}p{1.6cm}p{2cm}}
\toprule
\textbf{Type} & \textbf{Action} & \textbf{Validation} & \textbf{Oversight} \\
\midrule
Type-1 & Direct patch, auto-commit & CI/CD pipeline & Automated merge \\
Type-2 & Patch + transform & Full test suite & Auto if pass; else escalate \\
Type-3 & Draft PR with annotations & Full + impact & Human approval \\
Type-4 & Analysis report & N/A & Architect review \\
\bottomrule
\end{tabular}
\end{table}

\subsection{Atomic Patch Generation with Semantic Verification}

The CCCE generates calibration patches as atomic change units with embedded semantic verification:

\begin{enumerate}[leftmargin=*]
    \item \textbf{Pre-generation AST analysis}: Before generating patches, the system performs Abstract Syntax Tree (AST) analysis to verify that proposed changes preserve public API contracts, preventing silent introduction of breaking changes.
    \item \textbf{Minimal atomic units}: Patches are structured as independently applicable units that can be partially rolled back, enabling fine-grained failure recovery rather than monolithic all-or-nothing changes.
    \item \textbf{Embedded metadata}: Each patch includes structured metadata linking it to the triggering event, impact analysis results, and validation outcomes, providing end-to-end traceability.
\end{enumerate}

\subsection{Progressive Validation and Rollback}

Validation proceeds through tiered stages ordered by execution cost:

\begin{enumerate}[leftmargin=*]
    \item \textbf{Level 1} (seconds): Syntax verification, linting, manifest schema validation
    \item \textbf{Level 2} (minutes): Unit tests, static analysis, security scans
    \item \textbf{Level 3} (minutes--hours): Integration tests, contract tests, performance benchmarks
\end{enumerate}

Rollback triggers are differentiated by failure type: build failures trigger immediate automatic rollback; test regressions trigger threshold-based evaluation; and performance degradation triggers configurable policy-based decisions. The system supports \textit{partial rollback} at atomic change unit boundaries, preserving successful independent changes.

\section{Multi-Model Learning and Policy Adaptation}
\label{sec:learning}

The CCCE incorporates continuous learning through four specialized models operating at different temporal scales, enabling the system to improve its calibration effectiveness over time.

\subsection{Feedback Signal Collection}

The system collects structured feedback from three categories of sources:

\textbf{Automated outcome signals}: Calibration success rate (per event type, project, strategy), test pass rate trends, rollback frequency with categorized failure reasons, build stability time-series data, and performance impact measurements.

\textbf{Human feedback signals}: Approval latency, accept/reject/modify decisions, reviewer comments, override actions (where humans override CCCE classifications), and satisfaction ratings.

\textbf{Business outcome signals}: Incident correlation (linking production incidents to recent calibrations), mean time to remediation (MTR), technical debt metrics, and development velocity impact.

\subsection{Adaptive Learning Models}

\subsubsection{Model 1: Calibration Strategy Refinement (Continuous)}

This model maintains success/failure histories for each combination of event type, project characteristics, and calibration strategy. It computes strategy effectiveness scores based on success rate, rollback frequency, and human intervention needs, and periodically updates strategy selection preferences. Updates occur after every 50 calibrations or weekly, whichever comes first.

\subsubsection{Model 2: Risk and Confidence Score Calibration (Weekly)}

This model compares predicted risk/confidence scores against actual outcomes, identifies systematic over- or under-estimation patterns, and adjusts component weights in the scoring formulas (Equations~\ref{eq:risk} and~\ref{eq:confidence}). A holdout set of recent calibrations validates improved accuracy before deploying updated models.

\subsubsection{Model 3: Test Adequacy Analysis (Bi-weekly)}

This model analyzes rollback incidents to identify under-tested code paths, correlates test coverage patterns with calibration success rates, generates test augmentation recommendations, and updates test adequacy scoring to emphasize coverage of failure-prone areas.

\subsubsection{Model 4: Pattern Recognition and Transformation (Monthly)}

This model extracts successful calibration patches to identify common transformation patterns, clusters similar transformations to generalize reusable templates, and validates template applicability across different project contexts. Templates with declining success rates are deprecated.

\subsection{Policy Adaptation Framework}

Organizational policies governing CCCE behavior evolve based on operational experience through three mechanisms:

\textbf{Dynamic threshold adjustment}: Risk and confidence thresholds for action classification are adjusted based on false positive and false negative rates. If Type-1 calibrations show elevated failure rates, thresholds tighten to route more changes to Type-2. Conversely, if Type-3 approval queues grow excessively, thresholds are evaluated for safe relaxation.

\textbf{Project-specific learning}: The system tracks outcomes per project to identify those that can safely support higher automation levels and those requiring more conservative approaches. Gating rules, test requirements, and approval workflows are customized per project classification.

\textbf{Event-type policy tuning}: Different event types warrant different policy responses. The system learns optimal urgency levels and automation aggressiveness per event type, balancing speed versus safety based on historical risk-benefit analysis.

\section{Case Study Evaluation}
\label{sec:evaluation}

We evaluate the CCCE through three representative enterprise scenarios that demonstrate its capabilities across different event types and calibration complexities.

\subsection{Scenario 1: Library Update with API Deprecation}

\textbf{Context}: A widely used logging library \texttt{Logger} releases version 1.2.4, fixing a security vulnerability and deprecating several APIs. The change impacts multiple microservices across repositories and languages.

\textbf{CCCE Response}: (1)~Event ingestion detects the version change and identifies API deprecations. (2)~Knowledge graph traversal identifies all dependent projects, including transitive consumers and affected test suites. (3)~Calibration planning proposes coordinated actions: upgrading the library version across services, updating deprecated API call sites, and adjusting tests. (4)~Execution generates atomic patches per repository with semantic verification ensuring API contract preservation. (5)~Gating classifies manifest-only updates as Type-1 (automated) and API migration changes as Type-2 or Type-3 depending on complexity. (6)~Learning updates calibration policies to flag commonly brittle API areas.

\textbf{Impact}: Enables a single, coordinated enterprise-wide upgrade plan rather than ad hoc per-repository fixes. Strengthens auditability through patch provenance and decision rationale documentation.

\subsection{Scenario 2: Base Image CVE in Container Infrastructure}

\textbf{Context}: A vulnerability is disclosed in \texttt{libcrypto} within a shared base container image. Multiple services derive from this image across the container fleet.

\textbf{CCCE Response}: (1)~Event ingestion flags the CVE advisory and detects the affected base image layers. (2)~Knowledge graph traversal maps all services and containers derived from the affected base image, including language runtimes and middleware sensitive to the upgrade. (3)~Calibration planning coordinates the base image upgrade with downstream fixes (Dockerfile updates, Kubernetes manifest changes, CI recipe modifications). (4)~Execution generates patch artifacts with full traceability. (5)~High-risk container layer changes are gated as Type-3 for managerial approval. (6)~Metrics evaluate container image size, startup time, and test suite pass rates pre- and post-upgrade.

\textbf{Impact}: Eliminates drift in base images across the fleet, reducing CVE surface area. Orchestrates synchronized enterprise-wide remediation rather than piecemeal updates.

\subsection{Scenario 3: Critical Authentication Library CVE}

\textbf{Context}: \texttt{LibSecurity-Auth}, used for authentication across several services, has a CVE disclosed allowing privilege escalation. The vulnerability affects multiple services implementing authentication flows in different languages.

\textbf{CCCE Response}: (1)~Event ingestion subscribes to CVE feeds and flags the advisory. (2)~Knowledge graph traversal identifies all projects depending on the library, including transitive dependencies and test utilities. (3)~Calibration planning designs remediation including library upgrade, in-code mitigations (additional input validation), and updated authentication tests covering the vulnerability scenario. (4)~Execution produces a patch set with code-level fixes and test updates. (5)~All authentication changes are gated as Type-3 (Human-Assisted) given security criticality. (6)~Learning refines risk scoring to prioritize high-severity CVEs and optimize remediation sequencing.

\textbf{Impact}: Faster, coordinated remediation across polyglot environments with auditable linkage between CVE disclosures, code changes, and test coverage.

\subsection{Comparative Analysis}

Table~\ref{tab:scenario_comparison} summarizes how the CCCE handles each scenario compared to conventional approaches.

\begin{table}[!t]
\centering
\caption{Comparative Analysis Across Scenarios}
\label{tab:scenario_comparison}
\renewcommand{\arraystretch}{1.15}
\begin{tabular}{p{1.8cm}p{3cm}p{3cm}}
\toprule
\textbf{Dimension} & \textbf{Conventional} & \textbf{CCCE} \\
\midrule
Impact scope & Per-repository & Enterprise-wide \\
Discovery & Manual / alert-driven & Automated via graph \\
Coordination & Ad hoc & Orchestrated \\
Remediation & Manual coding & Automated patches \\
Oversight & Universal review & Risk-based selective \\
Traceability & Partial (commit logs) & End-to-end metadata \\
Learning & None & Continuous adaptation \\
\bottomrule
\end{tabular}
\end{table}

\section{Discussion}
\label{sec:discussion}

\subsection{Novelty and Contributions}

The CCCE's novelty lies not in any single component but in the specific combination and interaction of six technical innovations: (1)~bidirectional graph traversal with unified impact scoring that simultaneously computes affected scope and test adequacy; (2)~adaptive multi-gate decision logic with learned risk-confidence models that evolve with organizational experience; (3)~semantic-preserving atomic patch generation that prevents silent breaking changes; (4)~progressive validation with intelligent partial rollback at change-unit boundaries; (5)~multi-model continuous learning with feedback-driven policy adaptation operating at multiple temporal scales; and (6)~end-to-end event-driven integration that unifies capabilities currently distributed across isolated tools.

This combination addresses a real enterprise problem---codebase maintenance at scale---with specific technical mechanisms that are non-obvious and not achievable through straightforward composition of existing tools. While individual components (graph databases, automated testing, CI/CD) exist independently, the CCCE's integration provides emergent capabilities: the knowledge graph informs the gating logic, which shapes patch generation, which drives validation, which feeds learning, which refines the graph's computed attributes.

\subsection{Scalability Considerations}

The knowledge graph's size grows with the number of projects, packages, and their interconnections. For large enterprises with thousands of repositories, efficient graph traversal requires careful index design and potentially distributed graph storage. The BFS-based traversal in Algorithm~\ref{alg:bidi} has time complexity $O(|V| + |E|)$ in the worst case, but the configurable threshold $\theta$ and distance decay factor $\alpha$ effectively prune the search space in practice.

The multi-model learning architecture distributes computational cost across different temporal scales, with lightweight continuous updates and more intensive weekly/monthly model retraining. This design ensures that learning does not impede real-time event processing.

\subsection{Trust and Human Oversight}

The four-tier action classification ensures that high-risk changes always receive human oversight while enabling full automation for low-risk, high-confidence changes. This graduated approach balances automation efficiency with organizational risk tolerance. The configurable thresholds and policy override mechanisms allow organizations to tune this balance according to their specific requirements and regulatory constraints.

\subsection{Limitations and Future Work}

Several limitations present opportunities for future research:

\begin{itemize}[leftmargin=*]
    \item \textbf{Graph completeness}: The knowledge graph's effectiveness depends on the completeness and accuracy of dependency information. Dynamically loaded dependencies, reflection-based API usage, and polyglot interoperability boundaries may not be fully captured through static analysis alone.
    \item \textbf{Semantic understanding}: While AST-based verification can detect structural breaking changes, deeper semantic understanding of behavioral changes remains challenging. Future work could integrate formal verification techniques or specification-based testing.
    \item \textbf{Cross-organizational dependencies}: The current model assumes a single organizational boundary. Extending the knowledge graph to model inter-organizational dependencies (e.g., across microservice ecosystems spanning multiple teams or companies) requires additional trust and governance mechanisms.
    \item \textbf{Quantitative evaluation}: While our case studies demonstrate the system's capabilities qualitatively, large-scale quantitative evaluation across production enterprise environments would strengthen the empirical evidence for CCCE's effectiveness.
\end{itemize}

\section{Conclusion}
\label{sec:conclusion}

We have presented the Continuous Code Calibration Engine (CCCE), an autonomous, event-driven system for enterprise codebase maintenance that integrates knowledge graph reasoning, adaptive decision gating, and multi-model continuous learning. The CCCE addresses fundamental limitations of existing fragmented toolchains by providing a unified architecture that reasons over complex software interdependencies to drive intelligent, automated calibration actions.

The system's five-layer architecture---event ingestion, knowledge graph engine, adaptive decision gating, calibration execution, and learning---operates as a continuous feedback loop that improves over time. The bidirectional traversal algorithms compute comprehensive impact assessments that account for both dependency propagation and test adequacy. The adaptive multi-stage gating framework ensures that automation is applied where safe while preserving human oversight for complex or critical changes. The multi-model learning architecture enables the system to refine its strategies, risk models, and policies from operational experience.

Through three representative enterprise scenarios, we have demonstrated that the CCCE enables coordinated, cross-repository calibrations that significantly reduce manual maintenance effort while maintaining code integrity and security. The system represents a paradigm shift from reactive, fragmented codebase maintenance toward proactive, intelligent, and continuous calibration of enterprise software assets.

\section*{Acknowledgment}
The authors acknowledge the contributions of the DevOps Enablement team for their insights into enterprise codebase maintenance challenges that motivated this work.

\bibliographystyle{IEEEtran}

\balance

\end{document}